\documentclass[cits]{JINST}

\usepackage{amsmath, amsthm, amssymb}
\usepackage{mdwlist}
\usepackage{textcomp}
\usepackage{graphicx,subfigure}
\usepackage{fontenc}

\title{Optical Properties of Quantum-Dot-Doped Liquid Scintillators}

\author{C.~Aberle$^a$, J.J.~Li$^b$, S.~Weiss$^b$, and L.~Winslow$^a$\setcounter{footnote}{0}\thanks{corresponding author}\\
University of California, Los Angeles,\\
\llap{$^a$}Department of Physics \& Astronomy, 475 Portola Plaza, Los Angeles, CA 90095-1547, USA\\
\llap{$^b$}Department of Chemistry \& Biochemistry, 607 Charles E. Young Drive East, Los Angeles, CA 90095-1569, USA\\
  E-mail: \email{lwinslow@physics.ucla.edu}}

\abstract{Semiconductor nanoparticles (quantum dots) were studied in the context of liquid scintillator development for upcoming neutrino experiments. The unique optical and chemical properties of quantum dots are particularly promising for the use in neutrinoless double-beta decay experiments. Liquid scintillators for large scale neutrino detectors have to meet specific requirements which are reviewed, highlighting the peculiarities of quantum-dot-doping. In this paper, we report results on laboratory-scale measurements of the attenuation length and the fluorescence properties of three commercial quantum dot samples. The results include absorbance and emission stability measurements, improvement in transparency due to filtering of the quantum dot samples, precipitation tests to isolate the quantum dots from solution and energy transfer studies with quantum dots and the fluorophore PPO.}

\keywords{Scintillators, scintillation and light emission processes (solid, gas and liquid scintillators); Large detector systems for particle and astroparticle physics; Neutrino detectors}

\begin{document}
\section{Introduction}\label{intro}
Liquid scintillator detectors have been widely used in astroparticle physics over the last decades and were particularly successful in neutrino physics. Although much tuning of the optical properties has been done and new organic and organometallic compounds have been developed, most of the basic ingredients of scintillation-based detectors have remained unchanged. Progress in nanotechnology in recent years may hold the key to a new class of superior liquid scintillators which make use of the unique properties of quantum dots (QDs). In this paper, liquid scintillator requirements for neutrino detection are discussed in the context of QD doping and measurements of key optical properties of commercially available QD samples are presented. 

The QDs discussed here are small (typically 1.5-10~nm diameter) semiconductor nanocrystals in colloidal suspension. Quantum-confinement effects due to their small size lead to electrical and optical properties which are between the bulk material and single molecules. By changing the size or the composition of these crystals, the optical properties can be tuned \cite{alivisatos1996,swafford2006}. This behavior is highly interesting in the context of optimizing the energy transfers and the optical parameters of liquid scintillators. Narrow emission spectra tuned to match the quantum efficiency spectrum of suitable photodetectors (or absorption spectra of secondary wavelength shifters) can increase the quality of the scintillation signals, especially those coming from rare processes like neutrinoless double-beta decay (0$\nu\beta\beta$). 

The 0$\nu\beta\beta$ process is particularly interesting because, if observed, it would indicate that the neutrino is a Majorana particle (i.e. the neutrino is its own antiparticle). If this beyond-the-Standard-Model process exists it could be observed with several candidate isotopes \cite{vogel2002}. These isotopes have the property that single beta decay is energetically forbidden and thus can not contribute to the background. Different technologies are currently being explored to search for 0$\nu\beta\beta$, with liquid scintillator detectors among the possibilities \cite{kamlandzen,snoplus}. One major motivation to study quantum-dot-doped liquid scintillator is that commercial QDs can be made of 0$\nu\beta\beta$ candidate isotopes: $^{116}$Cd, $^{82}$Se, $^{128}$Te and $^{130}$Te. In fact, CdS and CdSe are among the most commonly used QD cores while CdTe is widely available as well.

In Section \ref{Requirements_section}, we discuss the requirements which have to be met by liquid scintillators used in particle physics experiments in general and neutrino experiments in particular. We also explain in more detail the motivations for the study of QD scintillators. Section \ref{Attenuation_length_section} contains the results of detailed attenuation length measurements of the components of the QD samples. Closely connected to the attenuation length measurements are the fluorescence properties of the QD solutions which are discussed in Section \ref{Fluorescence_section}. 

Properties of candidate QD samples have been previously reported \cite{mitpaper}. These results demonstrated that QDs are promising candidates which benefit further study. This paper is a continuation of this effort towards a complete characterization of quantum-dot-doped liquid scintillators.

\section{Requirements for Liquid Scintillators}\label{Requirements_section}
Several general requirements have to be met by liquid scintillators to be used in neutrino experiments. Although we focus on 0$\nu\beta\beta$, other applications of quantum-dot-doped liquid scintillators in neutrino physics are conceivable \cite{mitpaper} and many of the requirements listed here are more generally applicable. In general, rare earth element and metal-doped scintillators are interesting for a variety of future experiments. However, synthesizing stable doped scintillators is a difficult task \cite{gd_mpik,chooz,piepke}. Conventionally, single rare-earth or metal atoms are introduced in organometallic compounds. QDs dissolved in scintillators provide a genuine alternative to the conventional methods. In the following, we discuss scintillator requirements in the context of quantum-dot-doped scintillators.  

\paragraph{Solubility of the Components and Scintillator Stability}
One of the basic requirements is sufficient solubility of the scintillator components in the solvent material. QDs are readily available from commercial suppliers and they are typically dissolved in toluene at the level of several grams per liter. This is the order of magnitude needed to search for 0$\nu\beta\beta$ in a kiloton size detector. The QDs are coated with surfactants which bind to the surface of the QDs and allow the QDs to be dissolved in organic solvent. For neutrino experiments, typical run times are several years in order to accumulate the statistics needed to measure rare events. Therefore, the optical and chemical properties of QDs have to be stable over this time range. Stability tests in terms of absorption and emission have been carried out and are presented in Sections \ref{Attenuation_length_section} and \ref{Fluorescence_section}.

\paragraph{Safety and Material Compatibility}
Toluene is the standard solvent used by commercial suppliers to dissolve QDs. Although toluene provides a good starting point for the tests presented here, it may not be optimal for safety and material compatibility reasons in neutrino detectors. Toluene has a rather low flashpoint, 4~\textcelsius~compared to $>$20~\textcelsius~for other commonly used solvents \cite{yeh2007}, and it is known to be very chemically aggressive when in contact with frequently used detector materials like acrylics. However, the solvent can be exchanged by precipitation and re-dissolving of the QDs. Alternatively, companies can also provide QDs in powder form. 

\paragraph{Light Yield} 
The Light Yield (LY) of a liquid scintillator is directly related to the energy resolution of the detector system. A good energy resolution is crucial in 0$\nu\beta\beta$ experiments to reduce the background for this monoenergetic line signal. In particular, the tail of the standard model process of two-neutrino double-beta decay (2$\nu\beta\beta$) is an important irreducible background. It has been shown that the LY of quantum-dot-doped scintillators can be comparable to undoped scintillators\cite{mitpaper}. Further optimization of the composition is possible to increase the LY, for example a suitable additional wavelength-shifter could be added to minimize self-absorption losses.

\paragraph{Emission Spectrum}
The emission spectrum of a scintillator is matched to the quantum efficiency curve of the photodetectors in order to maximize the energy resolution. QDs provide a way to tune their emission spectrum by changing the size or the composition of the dots. The band gap in the semiconductor nanocrystals increases with decreasing size \cite{alivisatos1996}. In the measurements presented here, CdS core type QDs as well as alloyed CdS$_x$Se$_{1-x}$/ZnS have been used. In the alloyed QDs, the fraction $x$ of sulfur determines the emission wavelength at a constant size \cite{swafford2006}. 

The idea to do direction reconstruction of charged particles in liquid scintillators is described elsewhere \cite{mitpaper,simpaper}. Here, we note that QDs provide an interesting possibility of tuning the scintillation light such that the emission is at shorter wavelengths. The Cherenkov light, which is also produced by charged particles, has longer average wavelengths (if the spectra are weighted by the quantum efficiency of typical photodetectors and absorption effects are included). If this difference is increased, the arrival times of scintillation photons and Cherenkov photons will be separated to a larger extent due to the different speed of light for different wavelengths. Thus, the early Cherenkov light can be efficiently separated from the scintillation light by a time cut and the directionality of the Cherenkov light can be exploited. Because of this argument and in order to match typical photodetectors' responses, we studied QDs with the lowest available emission wavelengths. Experimental results on the emission spectra are presented in Section \ref{Fluorescence_section}. 

\paragraph{Transparency}
Competitive 0$\nu\beta\beta$ scintillator detectors typically have dimensions of several meters. This means that a highly transparent scintillator is needed with low absorbance in the wavelength range of the scintillator emission. Converting absorbance to attenuation lengths, neutrino detectors require attenuation lengths on the order of 10~m or longer. This paper contains absorption measurements on QD samples (see Section \ref{Attenuation_length_section}). The effect of additional filtering as well as QD precipitation tests are also presented (see Section \ref{precipitation_subsection}). In the literature there is a lack of precise QD absorbance spectra, particularly in the wavelength region above the main excitonic absorption peak. For applications on a smaller scale this is not relevant, but for neutrino experiments, it is crucial. Furthermore, at these low absorbance levels there can be batch-to-batch and time variations.

\paragraph{Timing}
The scintillator emission time profile affects the ability to do vertex reconstruction (and direction reconstruction) in a neutrino detector. It has been shown that this time profile is similar with and without QDs \cite{mitpaper}. 

\paragraph{Radiochemical Purity}
The purity requirements in terms of radioactivity for 0$\nu\beta\beta$ experiments are strict due to the large lifetimes and correspondingly low rates for 0$\nu\beta\beta$ decay. The current best limit on the half-life of 0$\nu\beta\beta$ decay is $t_{1/2}>1.7\cdot10^{23}$ years for $^{116}$Cd \cite{solotvina}. In particular, gamma backgrounds close to the Q value of the double-beta decay are to be minimized. First measurements of the gamma spectrum of a QD sample are currently in preparation and will be the subject of future work.   

\section{Attenuation Length Studies}\label{Attenuation_length_section}
This section describes a detailed analysis of the attenuation length in QD samples. The main goal is to identify suitable QD candidates for an experiment on the scale of several meters. In order to do this, we analyzed the absorbance in the wavelength region of interest above the main excitonic absorption peak where the scintillator emission and undisturbed Cherenkov emission will be. Three QD samples have been studied in this work. Table \ref{samples_table} summarizes their basic properties. All of the samples are prepared with a QD concentration of 5~mg/ml with toluene\footnote{CHROMASOLV\textsuperscript{\textregistered} Plus grade from Sigma Aldrich, $\geq$99.9\%.} as the solvent to obtain the concentrations given in the table. 

\begin{table}[th]
\caption{Quantum dot samples and basic properties as provided by the manufacturers \cite{nanoco, sigma, cytodiagnostics, crystalplex}. Two identical Lumidot CdS380 samples have been bought and named S1 and S2 in the text below. The numbers in the sample names give the approximate position of the main emission peak in nanometers (measurements of the spectra can be found in Section \protect\ref{Fluorescence_section}).}
  \begin{center}
    \begin{tabular}{llll}
      Name & CdS400 & Lumidot CdS380 (S1 and S2)& Trilite450 \\
      \hline\hline\\[-5px]
      Type & core-type CdS & core-type CdS & alloyed core/shell \\ 
           &               &               & CdS$_x$Se$_{1-x}$/ZnS \\
      Manufacturer/Vendor & Nanoco Tech. & Nanoco Tech./ & Crystalplex/ \\
                          &              & Sigma Aldrich & Cytodiagnostics \\
      Date of Purchase & October 2012 & February 2013 & May 2013 \\
      Ligands & Oleic Acid & Oleic Acid & Oleic Acid \\
      Concentration & 1.25~mg/ml & 1.25~mg/ml & 1.0~mg/ml \\
      Quantum Yield & 31~\% & ~30~\%  & ~50~\% \\
      Size & 1.8 to 2.3~nm & 1.6 to 1.8~nm & 5.5 to 6.5~nm\\[5px] 
      \hline
    \label{samples_table}
    \end{tabular}
  \end{center}
\end{table}

Part of the UV/Vis absorption measurements have been performed using a Shimadzu UV3101 PC photospectrometer. For this spectrometer a custom-made 10~cm cell holder has been used to increase the sensitivity. For other measurements, a Perkin Elmer Lambda25 UV/Vis spectrometer was employed. Both instruments work with a double-beam setup and have a deuterium UV lamp and a halogen lamp for white light. The Perkin Elmer Lambda25 spectrometer has a better intrinsic sensitivity which allows us to obtain accurate measurements with smaller samples of just a few milliliters. Additionally, this gave us the option to do valuable cross-checks with the two instruments. Since our measurements are close to the sensitivity limit of both instruments, the observed agreement of the cross-checks within the errors is important. The graph captions indicate which instrument and which cell length was used. Pure toluene has been taken as a baseline for all of the measurements in this section. 

\begin{figure}[tb]
\begin{center}
\subfigure[]{\includegraphics[scale=0.32]{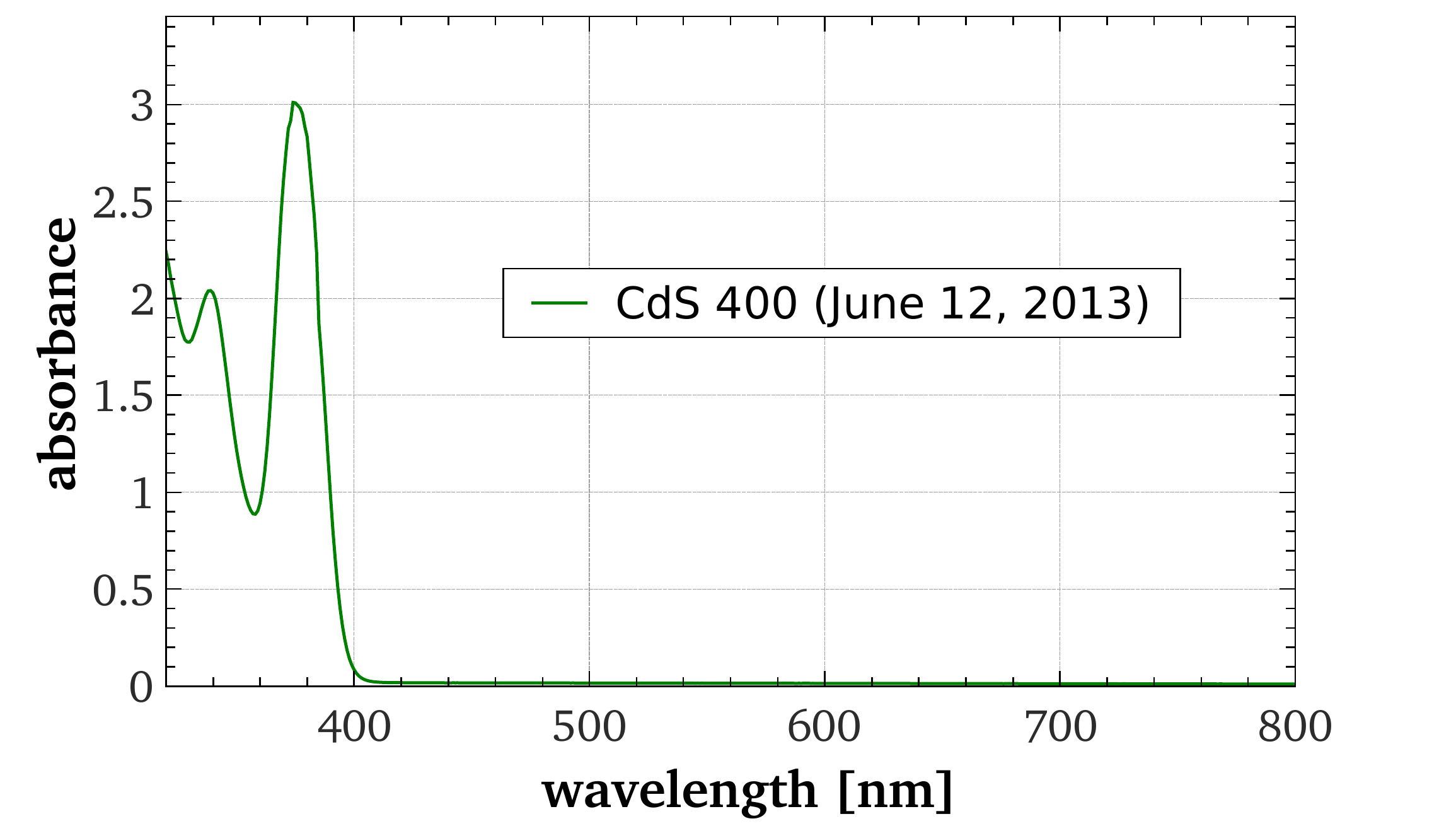}}
\subfigure[]{\includegraphics[scale=0.32]{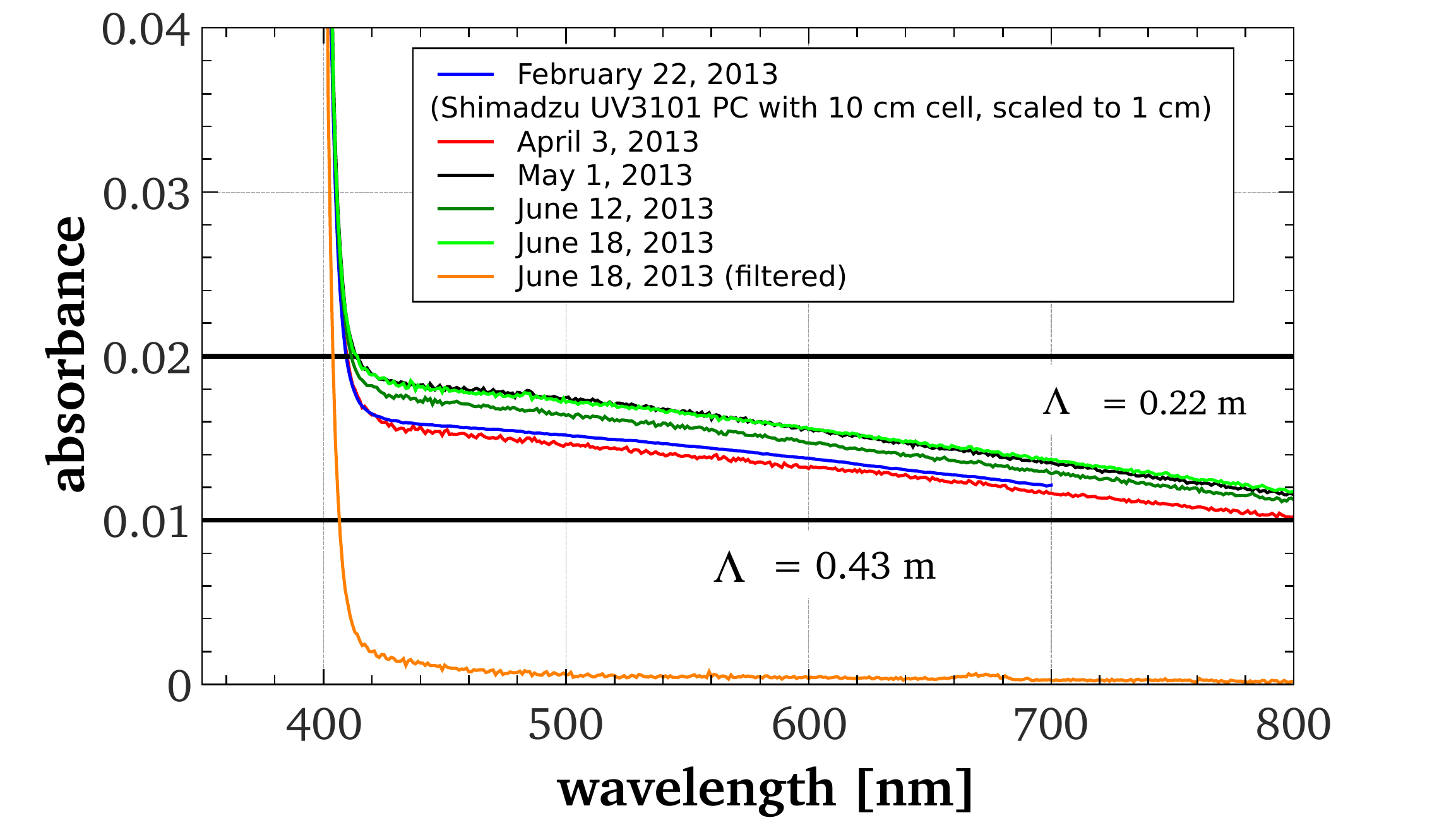}}
\caption[]{CdS400: (a) Absorbance of the CdS400 sample (1.25 mg/ml in toluene), as measured on June 12, 2013. (b) Zoom: The stability of the absorbance was tested on five different dates. The February 22 measurement was taken with the Shimadzu UV3101 PC instrument and a 10~cm cell. This measurement has been scaled by a factor of 0.1 to be compared to the other measurements. For all other data the Perkin Elmer Lambda25 instrument and a 1~cm cell was used. \label{cds400_stability}}
\end{center}
\end{figure}

\subsection{CdS400 Sample}
In Figure \ref{cds400_stability} (a) the absorbance of the CdS400 sample is shown. The main excitonic absorption peak can be clearly seen around 375~nm although the absorbance scale is not perfectly linear above about absorbance $A=1$. We focus on the quantitative analysis of the wavelength region above the main absorption peak, starting at $\sim$410~nm where the absorbance is well below $A=1$. This region is relevant when studying the transparency of the scintillator for its own emission spectrum (see Figure \ref{emission_comp} for the QD emission). 

From Figure \ref{cds400_stability} (b) we see that the absorbance is higher than 0.01 from 400~nm to 800~nm for the unfiltered CdS400 sample. The commonly used figure of merit in a meter-sized neutrino detector is the attenuation length
\begin{equation}
\Lambda = \textnormal{log}_{10}(e) \cdot \frac{x}{A(x)} \quad \textnormal{with} \quad A(x) = \textnormal{log}_{10} \left( \frac{I(0)}{I(x)} \right).   
\end{equation}  
The absorbance $A$ measured in a cell of length $x$ is the quantity shown in the figures of this section. The beam intensity before and after the cell has been passed is denoted with $I(0)$ and $I(x)$ respectively. Converting to attenuation length, we observe for the unfiltered CdS400 sample values smaller than 0.43~m below 800~nm. Filtering of the sample with a 0.2~$\mu$m PTFE membrane filter gives a significant improvement, a factor of $\sim$10, while the concentration of isolated QDs does not change (see also discussion below). The attenuation length is then above 2~m for wavelengths above 420~nm and the absorbance becomes too low to be accurately measured with the small cells. 
 
Another notable result is that the transparency of the CdS400 sample was not stable. From Figure \ref{cds400_stability} (b) we see that there is a trend towards higher absorbance values with time. Tests of the baseline stability and reproducibility of measurements lead to an error of $\Delta A = \pm 0.00075$ which is close to the manufacturer specification of $\pm 0.001$ at $A=1$ (Perkin Elmer Lambda25). For comparison, this is about the difference between the measurements taken on June 18, 2013 and June 12, 2013 at 500~nm. In contrast, the degradation between the first and last measurement (116 days apart) is significant at more than twice the baseline uncertainty. Explanations for the filtering and degradation effects and sample handling are discussed in Section \ref{discussion_sec} for all samples.  

\begin{figure}
        \begin{center}
        \subfigure[]{\includegraphics[scale=0.33]{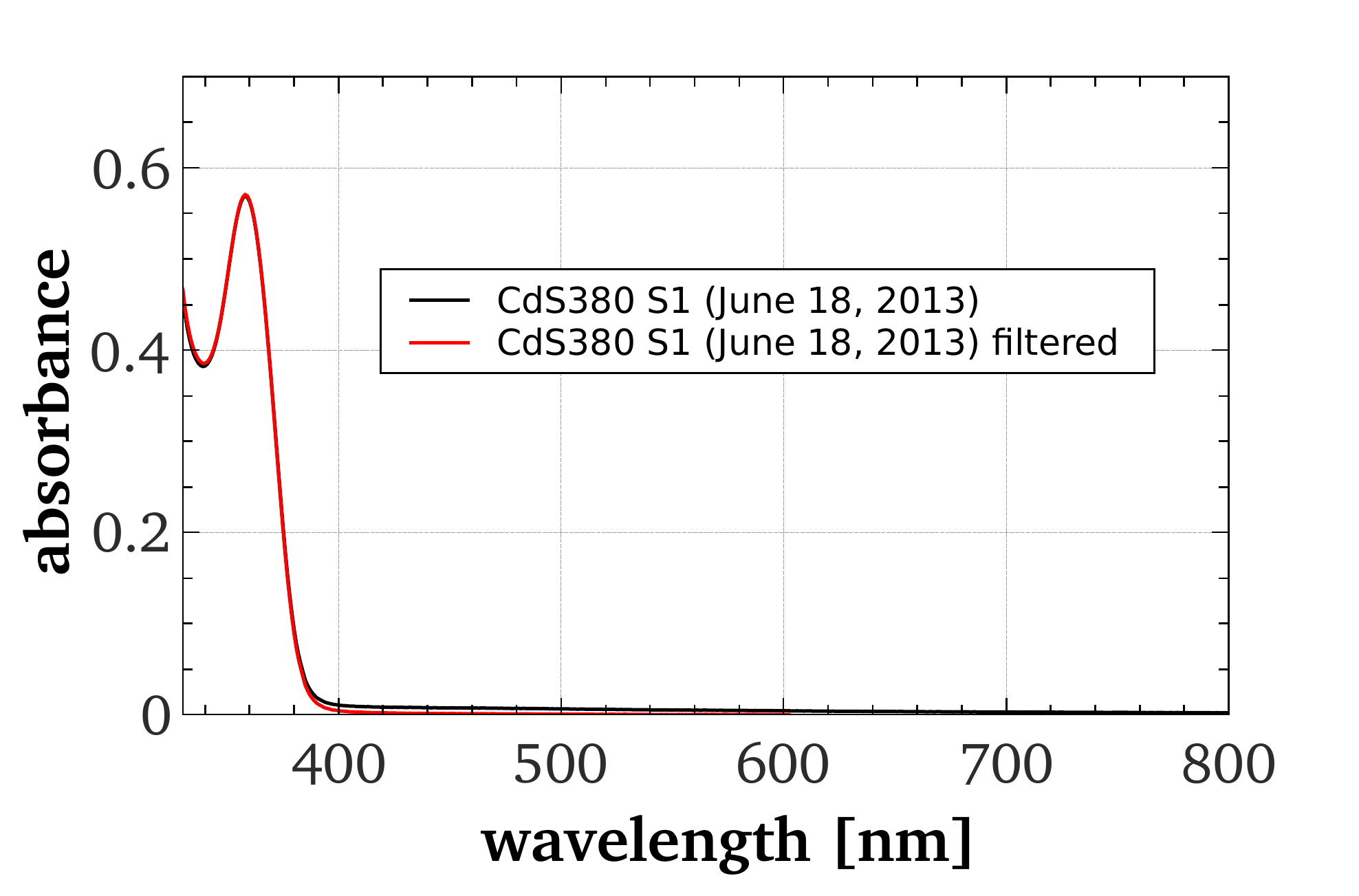}}
        \subfigure[]{\includegraphics[scale=0.34]{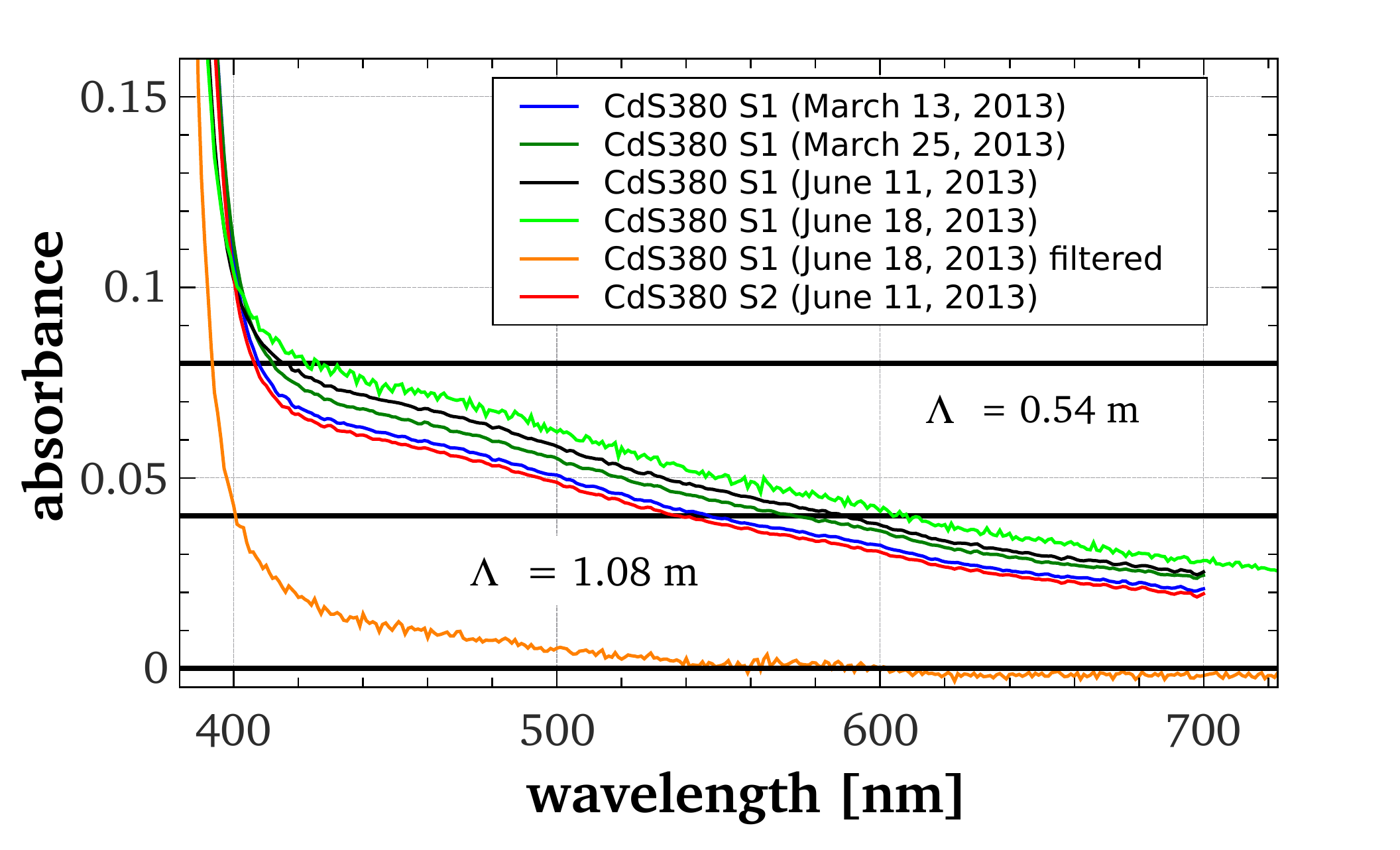}}
        \caption[]{CdS380: (a) Absorbance of the CdS380 S1 sample (1.25 mg/ml in toluene), as measured on June 18, 2013. The Perkin Elmer Lambda25 and a 1 cm cell have been used. (b) Absorbance of CdS380 samples over time and comparison with CdS380 S2 (stored in original, unopened bottle until June 11). The effect of filtering with a 0.2~$\mu$m PTFE filter is also shown. The 10~cm cell and the Shimadzu UV3101 PC instrument have been used except for the June 18, 2013 measurements where the Perkin Elmer Lambda25 and a 1 cm cell have been used. The measurements made on June 18 have been scaled by a factor of 10 to be compared to the other measurements. \label{cds380_stability}}
        \end{center}
\end{figure}

\subsection{CdS380 Sample}
Next, two identical CdS380 samples from Nanoco (bought via Sigma Aldrich), S1 and S2, are discussed (see Figure \ref{cds380_stability}). These samples have emission and absorption spectra which are on the low end of commercially available QDs. This is beneficial for direction reconstruction of charged particles in liquid scintillators (see discussion above and \cite{mitpaper,simpaper}). The attenuation length of the CdS380 samples are initially below one meter for wavelengths smaller than 550~nm. The values are higher than for the CdS400 sample but still not high enough for a large scale neutrino experiment with dimensions of several meters. 

We also observe that the transparency of the CdS380 S1 continually degrades over time. The difference after 90 days (March 13, 2013 to June 11, 2013) is $\Delta A = 0.008$ at 500~nm. The error of the Shimadzu UV3101 PC instrument has been estimated to be $\pm 0.002$ by multiple tests. This result is in close agreement with the manufacturer's specifications. Therefore, we conclude that the QD solution degradation is significant.
 
The two 10~ml samples of the same type (CdS380 S1 and S2) were used to do a comparative study of the transparency stability of the samples under different conditions. Sample S1 was opened on March 12, 2013 and then was repeatedly measured starting on March 13 while sample S2 was opened and first measured on June 11, 2013. The first measurements after opening the sealed containers of both S1 and S2 yield consistent results within the error of the instrument. This indicates that the sample is more stable in the sealed container and that the transparency degrades faster after opening the seal and with the handling of the sample.    

Filtering the CdS380 S1 sample improved the attenuation length significantly, returning to a value within instrumental sensitivity of the first measurement after opening. The QD concentration is largely unaffected by filtering since the main excitonic absorption peak, caused by monodisperse QDs, is unchanged. This is shown in Figure \ref{cds380_stability} (a).

\begin{figure}
        \begin{center}
        \subfigure[]{\includegraphics[scale=0.36]{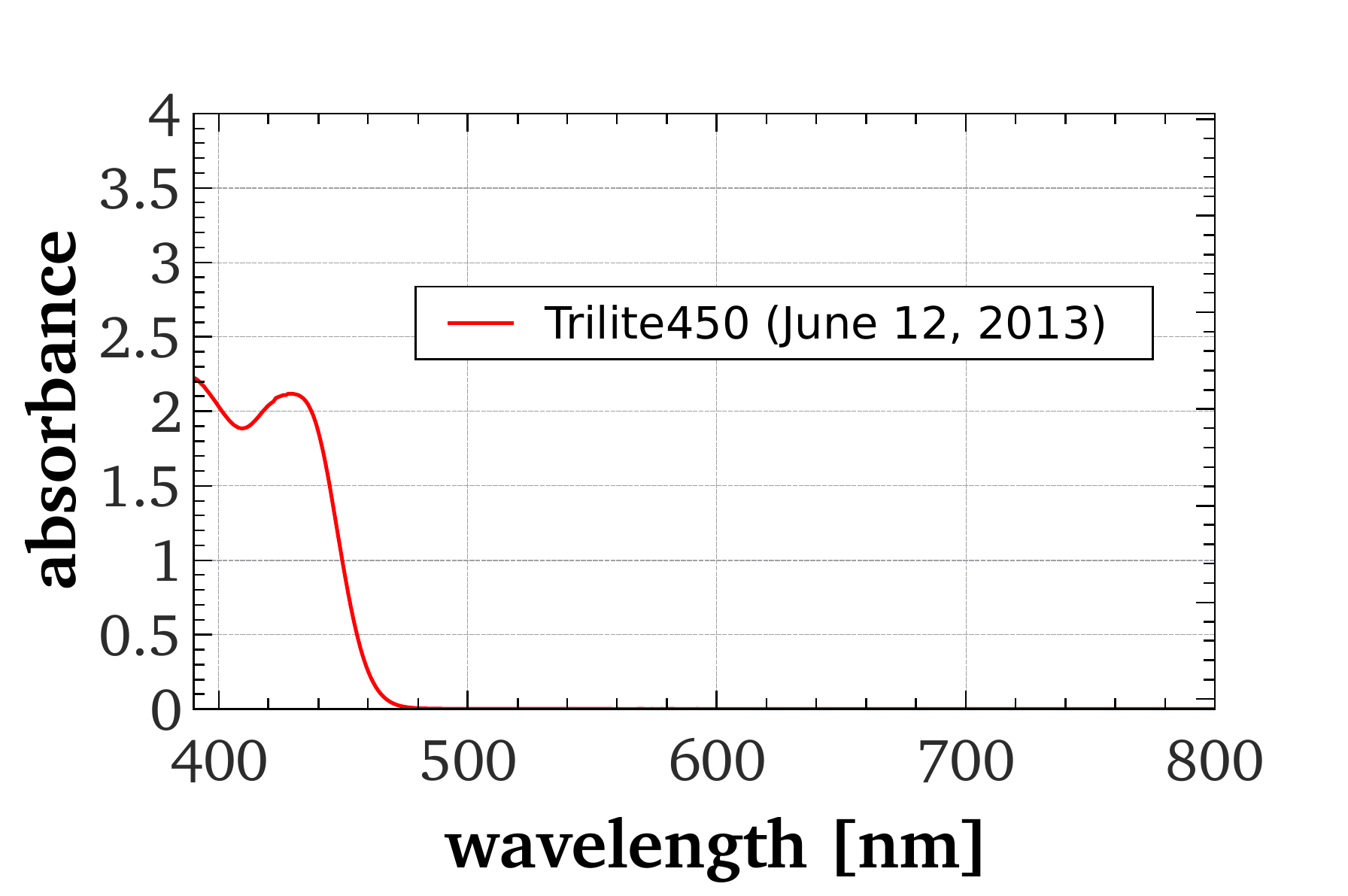}}
        \subfigure[]{\includegraphics[scale=0.36]{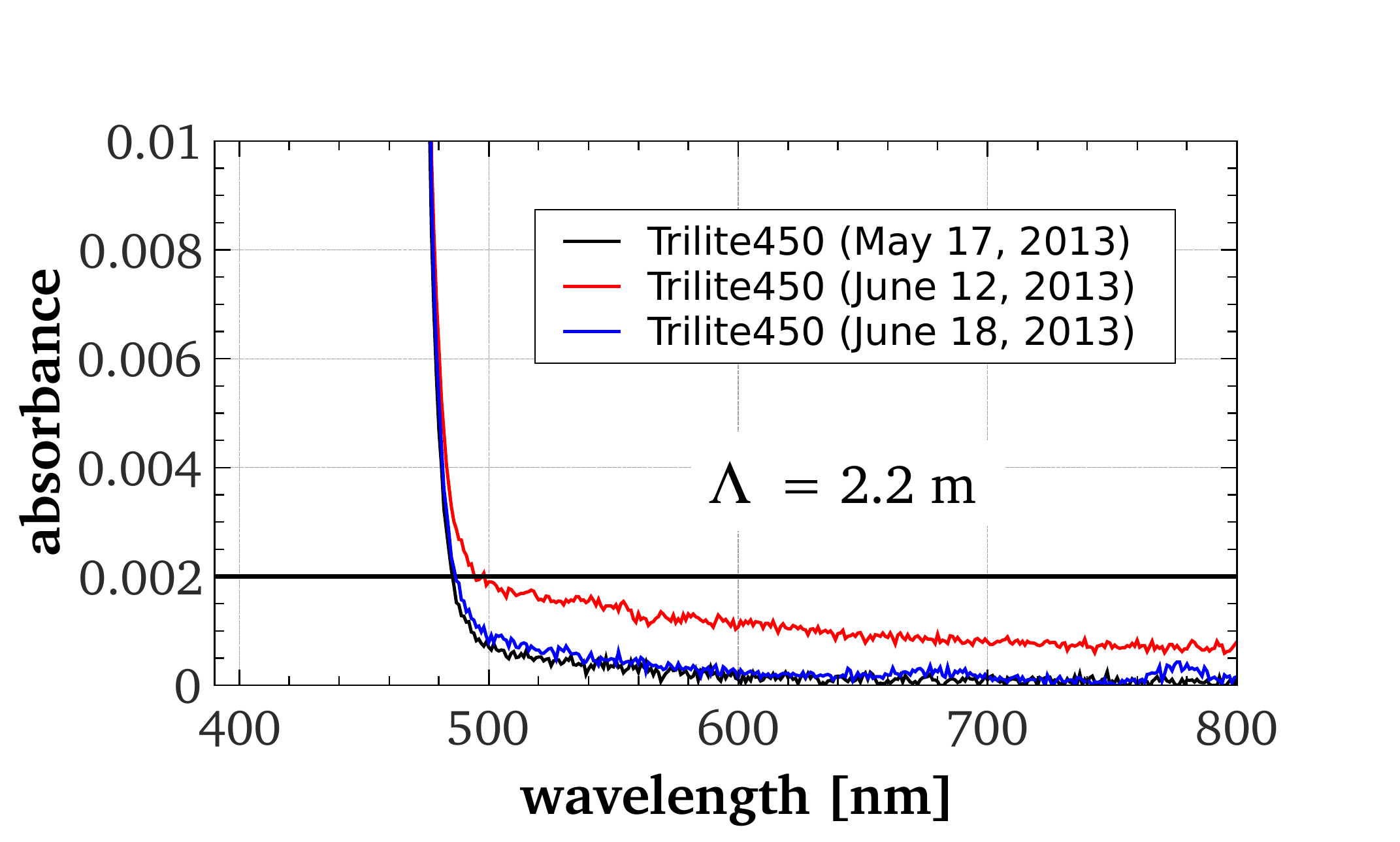}}
        \caption[]{(a) Absorbance of Trilite450 quantum dots in toluene (1~mg/ml). Note that in the peak region the absorbance becomes nonlinear (for A > 1). (b) Stability of the transparency over time for the Trilite450 quantum dot sample. The 1~cm cell and the Perkin Elmer Lambda25 instrument have been used for all the measurements in this graph. \label{trilite450_stability}}
        \end{center}
\end{figure}

\subsection{Trilite450 Sample}
A third sample was studied which, instead of the simple core type of the CdS400 and CdS380 QDs, has a more complicated structure. The Trilite450 QDs consist of a core (which is an alloy made of CdS$_x$Se$_{1-x}$) and a ZnS shell. The core-shell structure and the larger overall size of the QD should lead to more stable optical properties. The shell material ZnS has a higher band gap than CdS and CdSe, therefore excitons are confined to the core and do not overlap strongly with the surface trap states\cite{dabbousi1997}. This type of QD should also be less sensitive to degradation due to oxygen than the core-type QDs \cite{vanSark}. The composition of the core (the fraction $x$ of sulfur) determines the absorption and emission wavelengths while the size is kept constant. A further decrease in size, compared to the 5.5~nm to 6.5~nm for the studied sample, is conceivable in order to lower the emission and absorption wavelengths to those more appropriate for the proposed application. Smaller QDs are also known to have higher Stokes shifts \cite{swafford2006,capek,kuno1997} which helps to minimize self-absorption losses. At the time the sample was bought, 450~nm was the minimal wavelength available from the commercial suppliers we checked.

The results of the absorbance measurements for the Trilite450 sample are shown in Figure \ref{trilite450_stability}. The main excitonic absorption peak lies around 430~nm. There is a sharp drop in absorbance which ends at 485~nm. Above 485~nm, we find a low absorbance (see Figure \ref{trilite450_stability} (b)). The absorbance level corresponds to attenuation lengths of higher than 2.2~m above 500~nm. It should be noted that the error of the instrument is estimated to be $A = \pm 0.00075$ which is close to the absolute absorbance at 500~nm for the May 17 measurement. There seems to be a slight degradation in the transparency for this sample between the measurement on May 17 and June 12, 2013. When the sample was remeasured on June 18, 2013, the absorbance closely matched the original measurement. The absorbance increase between May 17 and June 12 and the decrease between June 12 and June 18 are covered by the error on the offset of $\pm 0.00075$ which matches the manufacturer specification. This means that in the future a more accurate measurement with a larger sample would be needed to determine the attenuation length of these promising QDs more precisely.
    
\subsection{Additional Measurements and Discussion}
\label{discussion_sec}
In order to study what causes the absorbance between 410~nm and 600~nm several analyses were done. First, individual known components of the QD solution were measured in terms of absorbance. As discussed above, filtering studies with the CdS solutions were also performed. Lastly precipitation with methanol and subsequent redissolving of the QDs was done to check the optical properties of the redissolved QDs. 

\paragraph{Absorbance of Ligands and Toluene}
Oleic acid is used as ligands in all of the studied samples to solubilize the QDs in toluene. One possible contributor to the $\mathcal{O}$(0.01) absorbance in Figure \ref{cds400_stability} is free, excess ligands. A sample of 10.0~mg/ml oleic acid (Aldrich, Technical Grade, 90\%) in toluene was prepared and the absorbance was the same with and without filtering of the sample. We used the same filter type which reduced the absorbance in the CdS380 S1 and CdS400 QD samples. Thus, we excluded that excess ligands in the QD samples cause the absorbance above the main excitonic absorption peak. 

Nanoco provided us with a sample of the same toluene used to prepare the CdS400 QD solution. UV/Vis measurements of this sample relative to the CHROMASOLV\textsuperscript{\textregistered} Plus toluene (used as a baseline for the UV/Vis measurements presented here) showed no significant absorption. The possibility that impurities in the toluene, which was used in the synthesis, cause the absorbance could thus be excluded.  

\begin{figure}
      \begin{center}
        \subfigure[]{\includegraphics[scale=0.34]{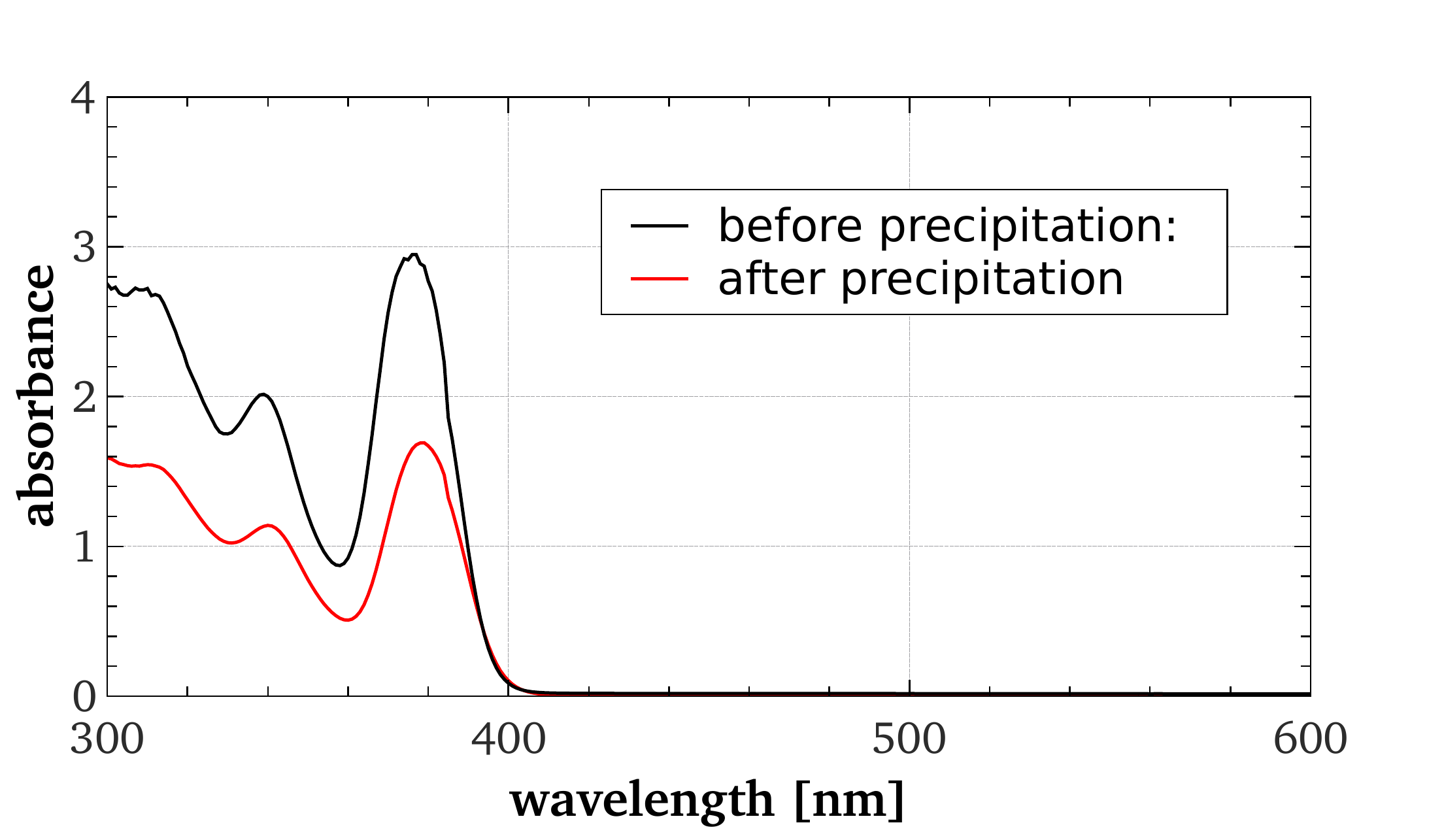}}
        \subfigure[]{\includegraphics[scale=0.32]{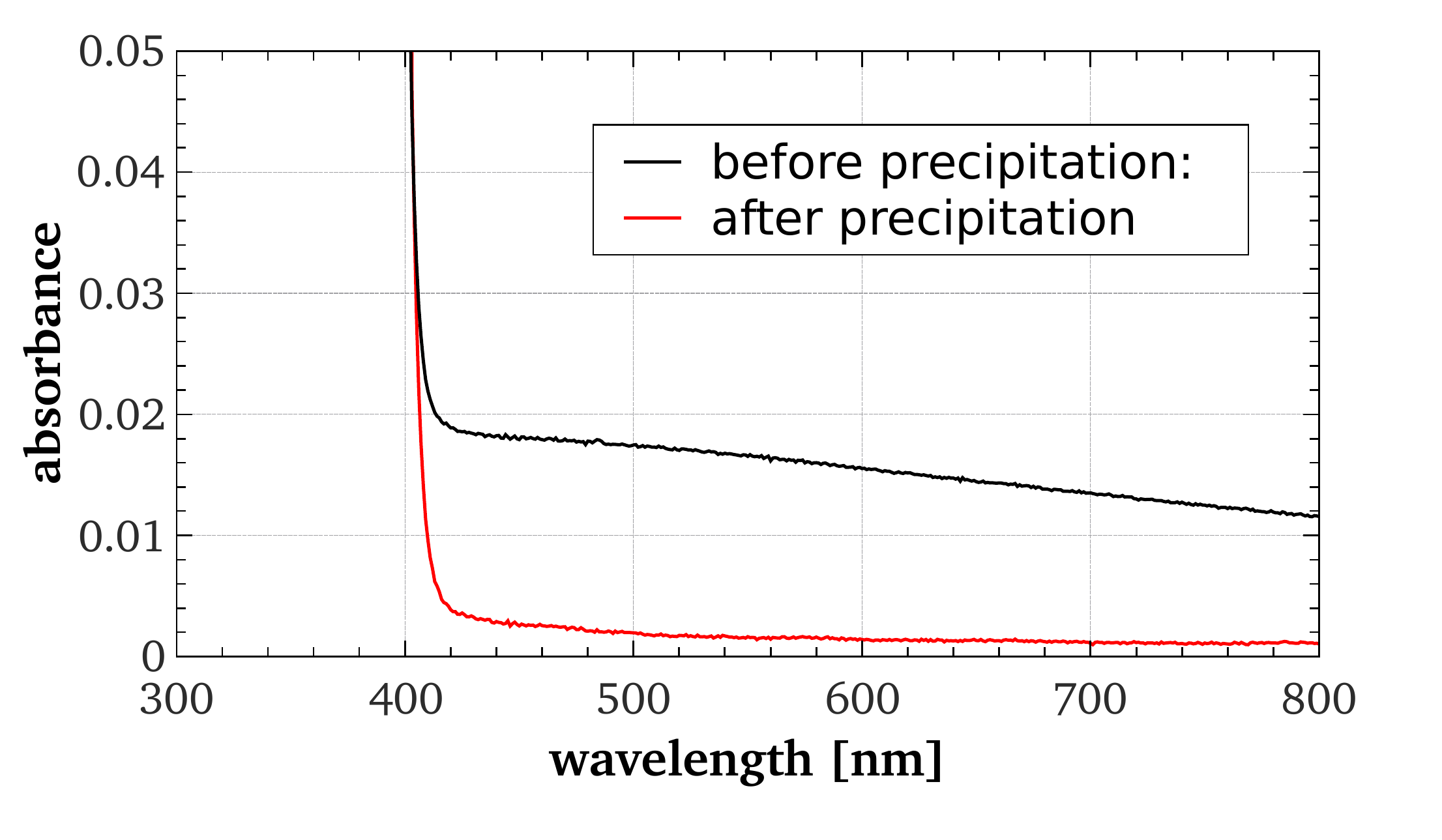}}
        \caption[]{Absorbance of the original CdS400 sample (1.25 mg/ml in toluene) compared to the absorbance after the precipitation. The measurements have been taken using the Perkin Elmer Lambda25 spectrometer and a 1~cm cell. \label{cds400_precipitation}}
        \end{center}
\end{figure}

\paragraph{Absorbance Improvement after Filtering and Sample Stability}
We saw that filtering of the original core-type QD solutions with 0.2~$\mu$m PTFE membrane filters resulted in significant improvement of the transparency. While the absorbance in the flat part of the spectra is reduced by factors of about 5 to 10 the absorbance in the peak region is stable. In addition, fluorescence spectra of the CdS380 S1 sample were measured before and after the filtering and the spectra are almost identical. For the CdS400 sample, the filtering causes only a small decrease in the fluorescence intensity, therefore the component which is filtered out is not the isolated QDs themselves. A possible explanation is that larger particles form by aggregation or self-assembly in the concentrated QD solutions over a long time scale. This could explain the increase in absorbance over time in this wavelength region and could simultaneously explain why this component can be filtered out. It would also explain the degradation in transparency over time. 

In addition, we have observed filaments in an unopened CdS core-type sample after storage at room temperature for 5 months. This sample was not used in these studies. The formation of large particles could explain the formation of filaments and the gel-like consistency of the initial 5~mg/ml CdS400 sample. The viscosity of the sample was high even after it was diluted from 5~mg/ml to 1.25~mg/ml. This indicates that for the Nanoco CdS core type QDs chemical stability may be an issue given the tight requirements of neutrino experiments. For reference, the manufacturers give a shelf life estimate of 1 year for all of the QDs. Finally, dust could contribute to the absorbance in the region of interest, and can be removed by filtering.  

Between the measurements, the samples have been kept in amber bottles and stored in the dark to protect them from UV light. Although the exposure to oxygen was tried to be kept minimal, it can not be excluded that oxygen plays a role in the transparency degradation of the sample. Samples were stored with a nitrogen blanket and the bottles were tightened carefully and sealed with Teflon tape and electrical tape. A possible contribution to the observed increase in absorbance for CdS380 and CdS400 is a QD concentration increase due to evaporation of toluene. Weight measurements were taken for some of the samples to test weight loss during the storage of the samples and no changes in weight were observed. The CdS400 sample had been stored at room temperature for 5 months and then at $\sim$8~\textcelsius~after opening. The CdS380 and Trilite450 samples were stored at $\sim$8~\textcelsius.

\paragraph{Precipitation of CdS400 Quantum Dots}
\label{precipitation_subsection} 
The last source of absorbance that was studied was residual compounds from the synthesis. We are lacking detailed information on these compounds because the synthesis is a proprietary process. However, we performed a precipitation of the CdS400 QDs with methanol: 1.5 ml of the original 1.25~mg/ml CdS400 sample were mixed with about 14~ml of methanol. A white, cohesive substance came out of solution and the remaining liquid became slightly turbid. After the initial precipitation step the sample was centrifuged to collect as many QDs as possible on the walls of the vial. Then, the supernatant was removed. The bottle with the white material and the QDs on the wall was dried with nitrogen and then the bottle was rinsed four times with toluene to redissolve the QDs. Thus, a 1.5~ml sample was prepared in order to compare with the original 1.25~mg/ml sample. As a last step, the sample was centrifuged to get rid of larger particles and dust. Figure \ref{cds400_precipitation} shows the comparison of the original sample with the sample after the precipitation. It can be seen that, in contrast to the filter treatment, the QD concentration was not fully recovered, indicating incomplete precipitation. However, the absorbance was disproportionally reduced in the region of interest. 

We wanted to estimate the recovery of precipitation by comparing the absorbance in a region where it is dominated by the QD absorbance but where the absorbance is still low enough to not be affected by the nonlinearity of the UV/Vis spectrometer. However, we observed a slight red shift of the spectrum in the peak region which made this approach less stringent. In order to study the cause of the shift in wavelength, we prepared a diluted CdS400 sample (dilution factor of about 10) and compared with the original sample. We did not find the strong peak shift which was obtained after the precipitation. This indicates that the precipitation step actually alters the photophysical properties of the QD sample. The best estimate of the fraction of QDs which are recovered after the precipitation is 58~\%. This has been obtained at the minimum around 380~nm where the absorbance is below 1 and the instrument is expected to still be close to linear. 

The precipitation step decreased the fluorescence emission intensity disproportionally (compared to the the peak region absorbance decrease) to about 30~\% of the original sample (not shown here). The effort to maximize the precipitation recovery likely caused surface ligand loss, possibly resulting in irreversible flocculation in a small portion. The small population of aggregated QDs in toluene could increase the apparent average size of the particles and cause a slight red shift of the peaks. Additionally, shifts in the absorbance spectrum have been described depending on the passivation of the QD surface with ligands and coordination of surface atoms (\cite{inerbaev} for CdSe QDs). The decrease of fluorescence intensity is also a result of surface ligand loss in the precipitation. 

\section{Fluorescence Spectra Measurements}\label{Fluorescence_section}
One of the main advantages of using QDs in liquid scintillators is that the emission wavelength can be controlled. Selecting QDs with a sharp emission spectrum allows more precise tuning of the emission with respect to the photodetector quantum efficiency or tuning to optimize the energy transfer to a secondary wavelength shifter. A sharp emission spectra also aids the separation of Cherenkov light from scintillation light for direction reconstruction. In this section, we present the fluorescence properties of the three samples listed in Table \ref{samples_table}. 

\begin{figure}
      \begin{center}
        \includegraphics[scale=0.42]{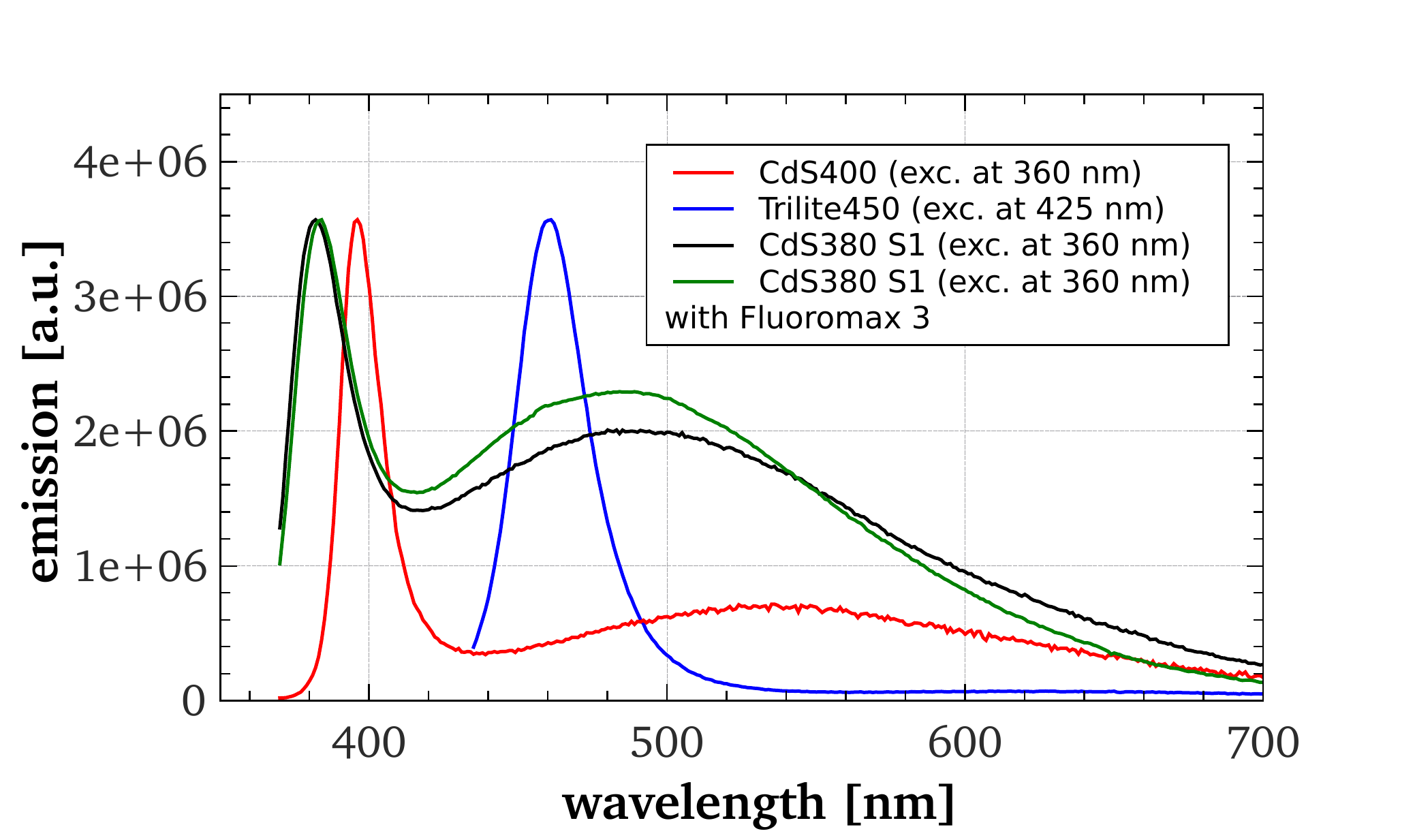}
        \caption[]{Fluorescence emission spectra of the three QD samples (normalized at the peak). A Photon Technologies International (PTI) QM-6SE instrument has been used. The CdS380 S1 sample was tested additionally with a second spectrometer (Horiba FluoroMax3) to verify the strong emission tail. All the spectra were recorded in the right angle (RA) geometry with a 1~cm $\times$ 1~cm UV transparent cell and corrected for the instrument response for better comparability. Photodetector dark count subtraction was not needed since the emission intensities were high. \label{emission_comp}}
        \end{center}
\end{figure}

A comparison of the emission spectra of the pure samples is shown in Figure \ref{emission_comp}. Both CdS400 and CdS380 show strong emission signals at wavelengths above the main narrow band gap emission line. These broad emission peaks come from surface trap state emission. The smaller CdS380 QDs show relatively more surface trap state emission compared to CdS400, which is expected \cite{capek}. In contrast, the Trilite450 sample does not show surface trap state emission but only the band gap emission around 461~nm. This difference can be explained by the effect of the passivating ZnS layer around the CdS$_x$Se$_{1-x}$ core. The ZnS shell provides an effective energy barrier and prevents photo-excited carriers from leaking through to the particle surface \cite{dabbousi1997}. The shape of the emission spectrum of the Trilite450 sample is preferred.

\subsection{Degradation with Intense Light Exposure}
\begin{figure}[tbh]
      \begin{center}
        \includegraphics[scale=0.39]{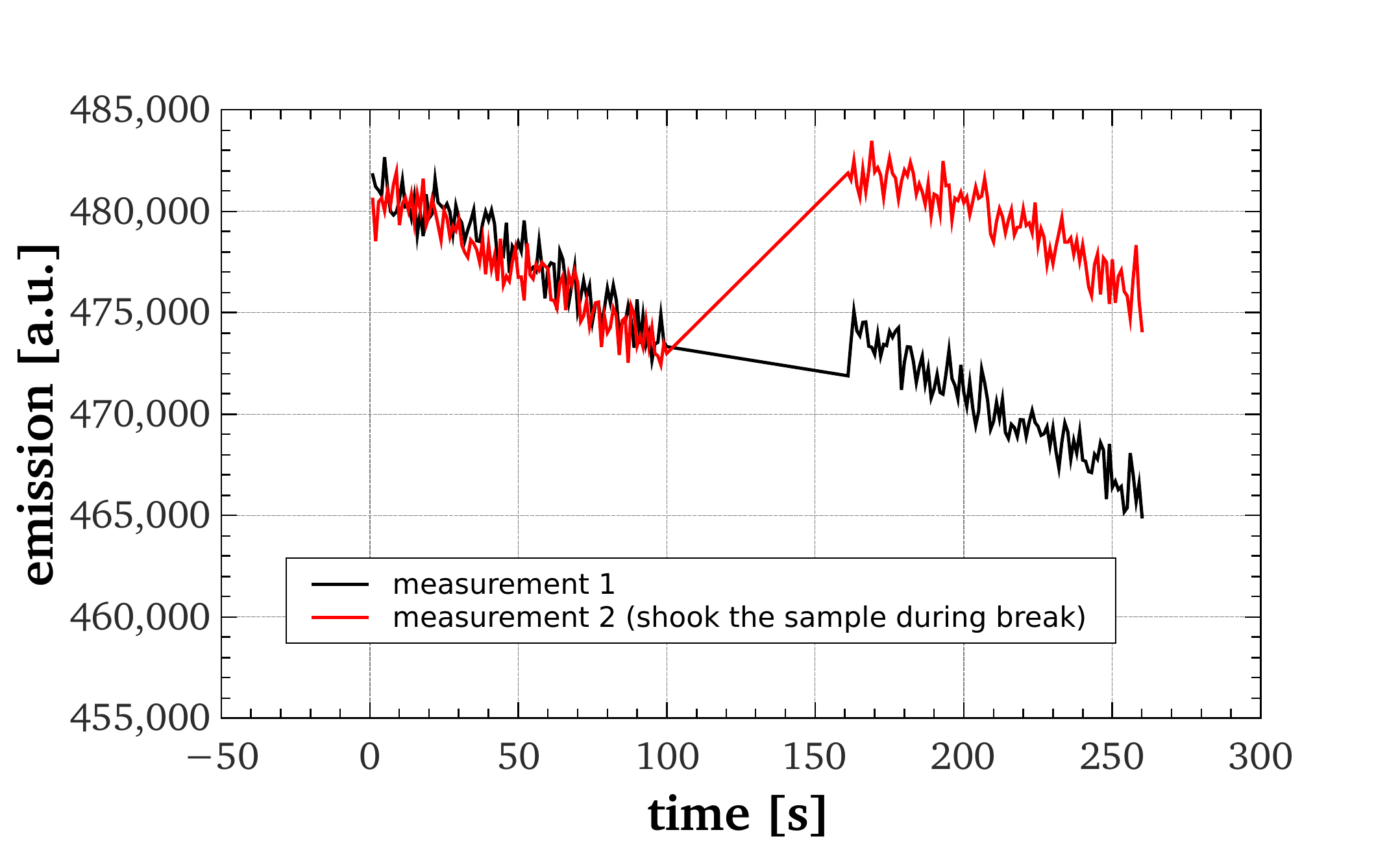}
        \caption[]{Time-lapse scan of the emission intensity at a fixed wavelength of 395.8~nm and a fixed excitation wavelength of 360~nm. The sample studied is the CdS400 sample and the PTI QM-6SE instrument is used. Between 100 and 160 seconds there is a break (the shutters are closed) which can be used to manipulate the sample. \label{time_scan}}
        \end{center}
\end{figure}

We observed that irradiation of the samples during the fluorescence measurements has an adverse effect on the emission intensity. Figure \ref{time_scan} shows for the CdS400 sample that the emission decreases with time during the excitation periods of a time-lapse scan measurement. When the instrument's shutter is closed the emission intensity stays at the same level and continues to drop once the scan resumes (black line). If the sample is shaken during the break, the original emission intensity is recovered (red line). Also turning the cell by 90\textdegree ~restores the initial intensity. All three samples show the decrease in fluorescence emission over time when excited at their main absorption peak while recording at an emission wavelength fixed to the main band gap emission peak. Further tests have been done recording complete spectra repeatedly. The ratio between the main band gap emission line and the surface trap state emission decreases for the CdS380 S1 and the CdS400 samples. 

The degradation effect can be explained by saturation or photodegradation of the QDs in a small region of the sample. The beam hits the sample always at the same point and since the solutions are rather viscous the QDs do not diffuse away from this region easily. The relative emission intensity loss was also measured for the two other samples. The effect was larger for the CdS380 S1 sample (compared to CdS400) and smaller for the Trilite450 sample. We note that it is difficult to compare the degradation of the QD samples quantitatively because there are differences in excitation beam intensities for the different samples due to different excitation wavelengths. In the literature, the process of photoinduced oxidation and bleaching of QD emission has been discussed in detail for single CdSe QDs \cite{vanSark}. It is expected that this effect is more pronounced for the CdS380 and CdS400 samples because they have no shell. Although liquid scintillators in the context of neutrino detectors are typically handled under nitrogen atmosphere and with little exposure to (UV) light, it is desirable to use optically stable components. From this point of view, the Trilite450 sample is preferred. Apart from the short-time decrease due to the fluorometer excitation no larger degradation effects were found on longer time scales. The emission spectra of CdS380 S1 (stored protected from UV light) taken 10 days apart in time were almost identical.

\subsection{Energy Transfer in Quantum-Dot-Doped Scintillators}
In this section, we present results on the energy transfer in the three QD samples with and without the commonly used wavelength-shifting fluorophore PPO (2,5-Diphenyloxazole). In liquid scintillators, charged particles predominantly excite the solvent molecules (toluene in our case) due to their high abundance in the mixture. Efficient (high rate) energy transfer from the solvent molecules to the solutes (PPO and/or QDs) is crucial for a high light yield because the direct emission of light from toluene undergoes strong self-absorption and subsequent light loss. After energy is transferred to the solutes via emission and absorption of real photons or by nonradiative energy transfer \cite{foerster}, the solutes can emit light at higher wavelengths where the scintillator is more transparent. With a fluorometer, the toluene molecules can be selectively excited at 260~nm and energy transfer can be directly studied. The observed emission spectra convey spectral information about the efficiency of the energy transfer paths to QDs and PPO. In light yield measurements \cite{mitpaper} the samples are excited with charged particles. These measurements can be compared qualitatively to the results presented here.

\begin{figure}
      \begin{center}
        \includegraphics[scale=0.44]{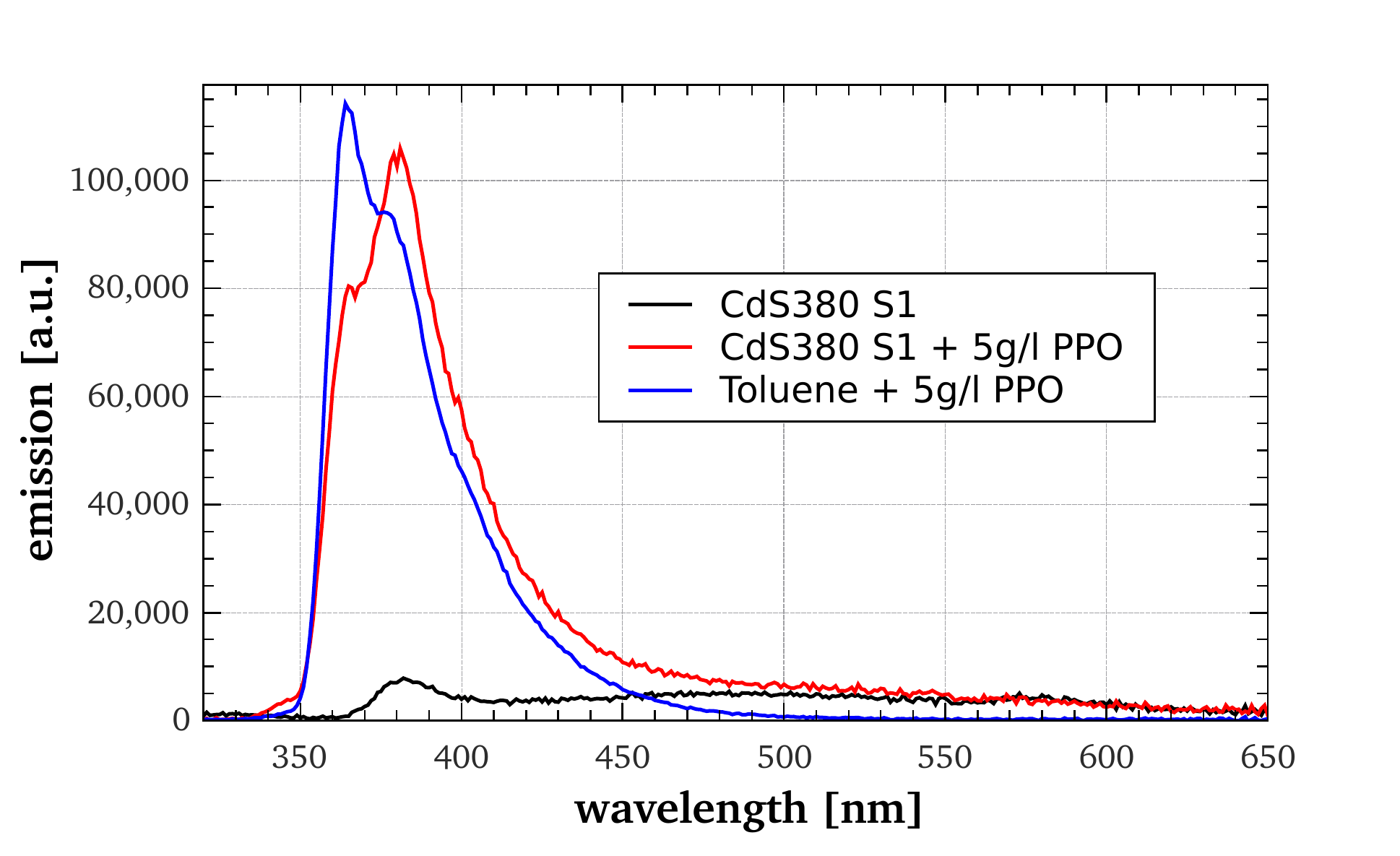}
        \caption[]{Comparison of the emission for the CdS380 S1 sample, CdS380 S1 + 5~g/l PPO and toluene + 5~g/l PPO. The excitation wavelength is 260~nm. For all samples the dark count rate (625 counts) of the instrument (PTI QM-6SE) has been subtracted and the wavelength-dependent detection efficiency correction has been applied. \label{G_QD380_ex260_corr}}
        \end{center}
\end{figure}

Figure \ref{G_QD380_ex260_corr} shows the emission of the toluene-based CdS380 S1 sample with and without the addition of 5~g/l PPO. For comparison, the emission spectrum of toluene plus 5~g/l PPO (without the QDs) is shown. The emission intensity of the CdS380 S1 sample without PPO is lowest. This indicates that direct energy transfer from toluene to the QDs at concentrations of 1.25~g/l is not very effective. Toluene plus PPO gives a higher light output. We find that the emission spectrum of the CdS380 S1 + 5~g/l PPO sample is a combination of the toluene + 5~g/l PPO spectrum and the CdS380 S1 spectrum. Due to the presence of the QDs, the PPO emission peak at 365~nm is weaker while the emission around 380~nm is stronger.  

\begin{figure}
      \begin{center}
\subfigure[]{\includegraphics[scale=0.36]{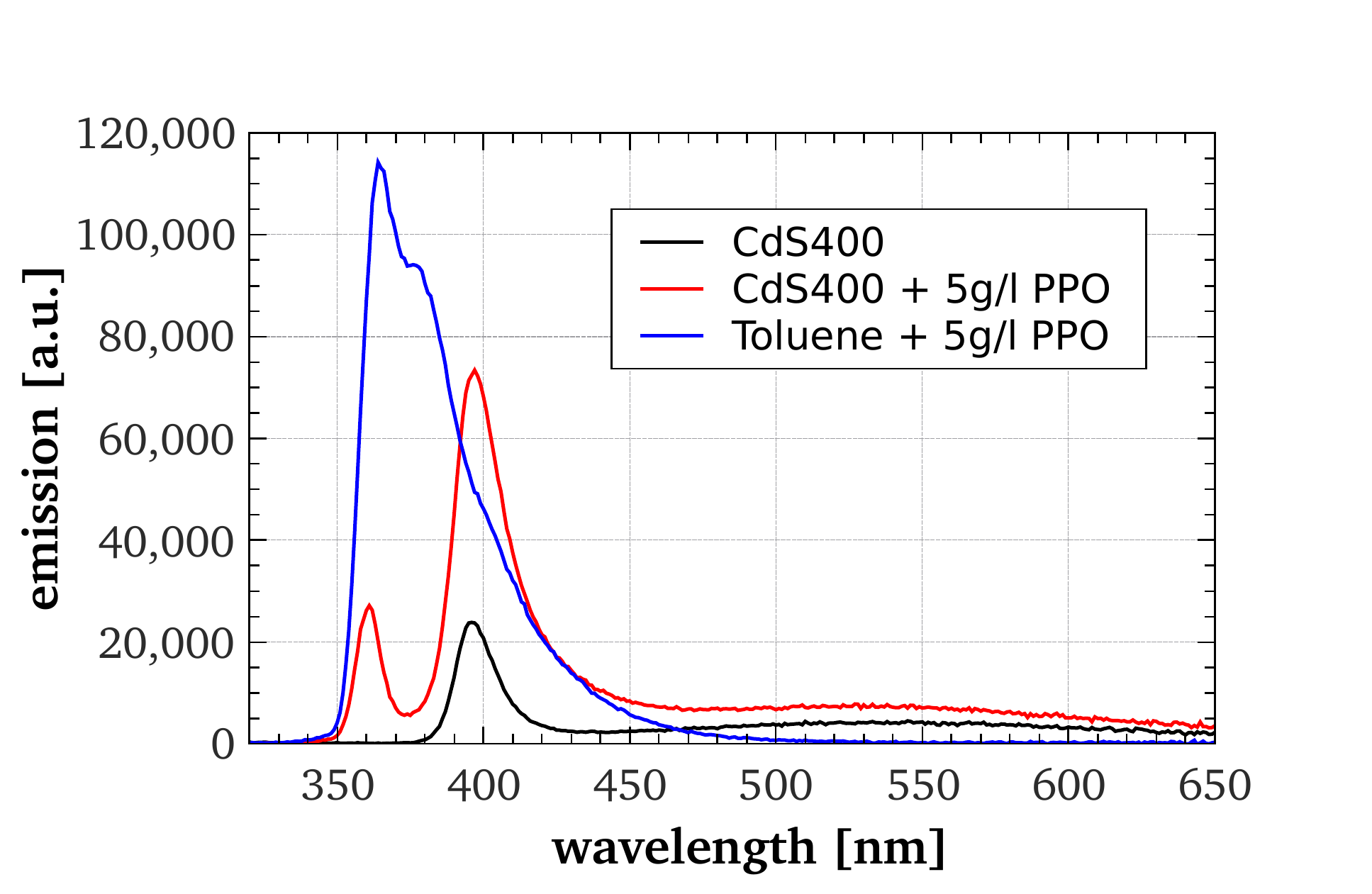}}
\subfigure[]{\includegraphics[scale=0.36]{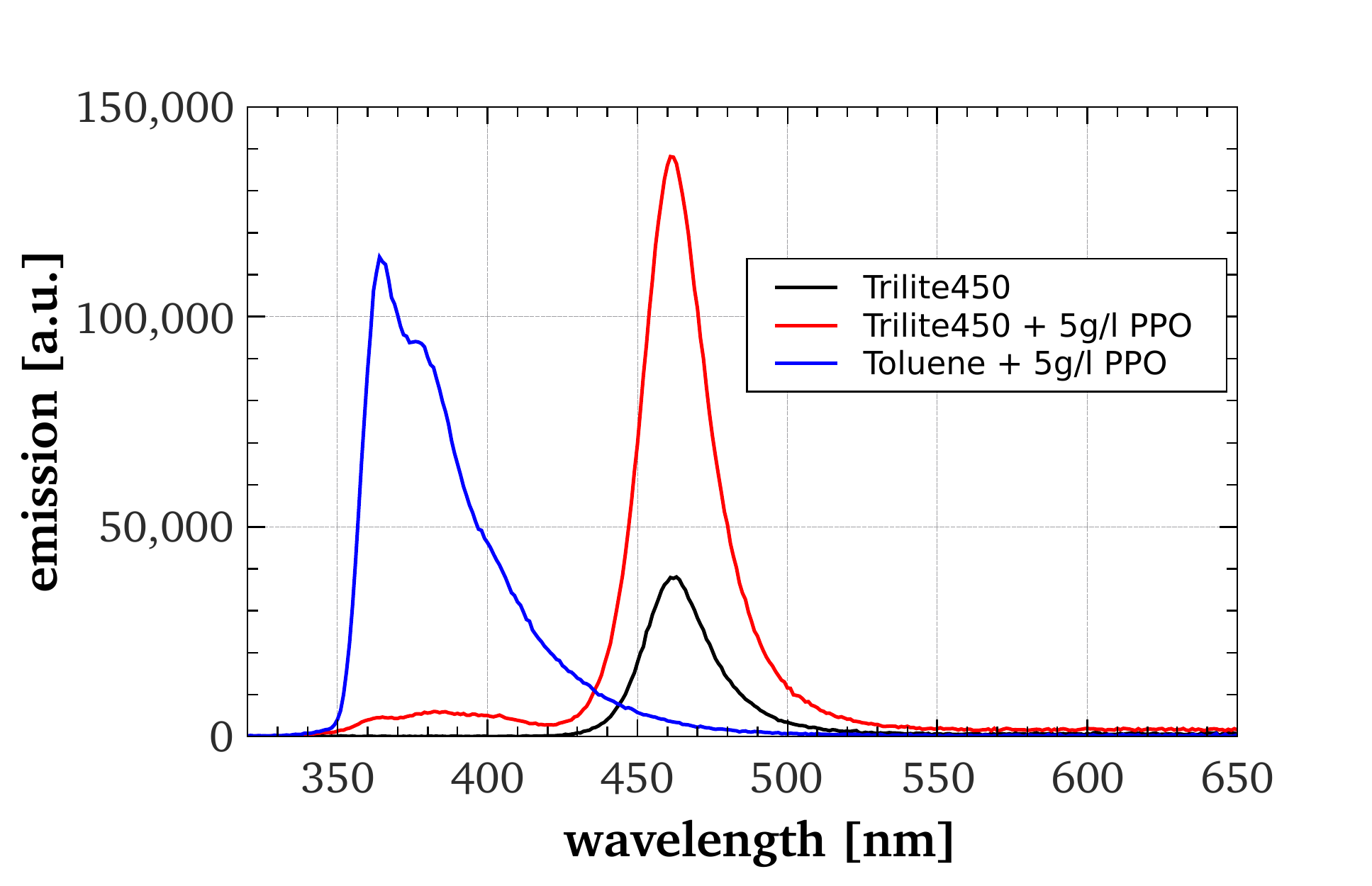}}
\caption[]{(a) Comparison of the emission for the CdS400 sample, CdS400 + 5~g/l PPO and toluene + 5~g/l PPO. (b) Trilite450, Trilite450 + 5~g/l PPO and toluene + 5~g/l PPO emission spectra. (a) and (b): The excitation wavelength is 260~nm. For all samples the dark count rate (625 counts) of the instrument (PTI QM-6SE) has been subtracted and the wavelength-dependent detection efficiency correction has been applied. \label{G_QD400_QD450_ex260_corr}}
        \end{center}
\end{figure}

In Figure \ref{G_QD400_QD450_ex260_corr} the same sets of data are displayed for CdS400 and Trilite450. For CdS400, we see that with the addition of QDs, the PPO emission is suppressed while the QD emission increases beyond the QD sample alone. Thus, we observe energy transfer from PPO to the QDs. For the Trilite450 QDs, the QD emission intensity is strongly enhanced by adding PPO while the PPO emission is strongly suppressed.  This indicates even more efficient energy transfer from PPO to the Trilite450 QDs compared to CdS400 and CdS380 due to high overlap of the PPO emission and QD absorption spectra \cite{foerster}.  

Light yield estimates are now calculated for the three PPO-loaded QD samples relative to toluene + 5~g/l PPO. The corrected fluorometer emission spectra discussed above are multiplied with the quantum efficiency spectrum of a Hamamatsu R1828-01 photomultiplier tube used in \cite{mitpaper} for light yield measurements. Subsequent integration of the spectra gives numbers which are proportional to the light yield or, more accurately, the photoelectron yield. The results relative to the undoped toluene + 5~g/l PPO scintillator are 112~\% (CdS380 S1 + 5~g/l PPO), 66~\% (CdS400 + 5~g/l PPO) and 76~\% (Trilite450 + 5~g/l PPO). Light yield measurements of 20~ml samples of CdS core type QDs using excitation from a radioactive $^{90}$Sr source have been reported earlier \cite{mitpaper}. Differences in self-absorption effects between samples resulting from different experimental geometries preclude a direct comparison of relative light yield, however smaller CdS QDs showed higher light yield in the previous study, which is consistent with our results. 
  
Since the quantum yield of the QD samples studied here is on the order of 30-50~\% (see Table \ref{samples_table}), too many reabsorption processes will result in photon loss with high probability. Core shell QDs, like the Trilite450, are known to have generally higher quantum yields than core-type QDs \cite{swafford2006,dabbousi1997,obrien2011} and quantum yields can still be improved, for example by better control of shell growth \cite{mcbride2006}. The Stokes shift of the studied QDs may also be too small to use them as the only solute in large scale liquid scintillator systems due to self-absorption. This implies that the use of secondary wavelength shifters together with the QDs might be necessary depending on the size of the detector system. The study of suitable wavelength-shifters or alternatively the study of QDs with higher Stokes shift would be the subject of future work. 

\section{Conclusions}

QDs are interesting candidates as components in novel liquid scintillators due to their unique optical and chemical properties. They provide an alternative way to dissolve isotopes and simultaneously enhance the fluorescence properties of liquid scintillators. Nevertheless, the multiple requirements for liquid scintillators used in neutrino detectors require thorough study of the QD properties. This work contributes to a more complete characterization of QD samples for the application in liquid scintillator detectors. 

It has been shown that attenuation lengths on the order of several meters for wavelengths above the main excitonic absorption peak are possible. Filtering with 0.2~$\mu$m PTFE filters has been found to improve the transparency of the CdS380 and CdS400 samples and might be considered as a purification step. The Trilite450 sample already showed high transparency without filtering. The energy transfer from toluene to the QDs, and its enhancement by the addition of the fluorophore PPO, was demonstrated. The alloyed core-shell Trilite450 QDs have higher quantum yield, are chemically and photophysically more stable and have a preferable emission spectrum shape compared to the core-type QDs. On the other hand, smaller QDs like CdS380 and CdS400 have higher Stokes shift and applications like direction reconstruction in liquid scintillators benefit from their lower emission wavelengths. 

Open questions include how the effect of self-absorption can be minimized and if the stability of QDs is sufficient over several years. As technology advances concerning the production of new QD types, high quantum yield, low overlap between absorption and emission spectrum (for example due to better size control) and high stability due to improved surface passivation for small size QDs would further improve the prospects of quantum-dot-doped liquid scintillators.

\acknowledgments
The authors would like to thank Christopher Coy for preliminary measurements on this topic. The authors would like to thank Ignacio Martini for his help with the UV/Vis measurements and useful discussion. The authors would also like to thank Prof. Benjamin Schwartz' lab for help with the fluorometer measurements. S. Weiss acknowledges the financial support through NIH grants 5R01EB000312 and 1R01GM086197. C. Aberle and L. Winslow are supported by funds from the University of California Los Angeles.

\end{document}